\documentclass[conference]{IEEEtran}
\usepackage{cite}
\usepackage{amsmath,amssymb,amsfonts}
\usepackage{algorithmic}
\usepackage{graphicx}
\usepackage{textcomp}
\usepackage{xcolor}
\usepackage{enumerate}
\newcommand{\mr}{\mathrm}                   
\newcommand{\mc}{\mathcal}                  

\newcommand{\veg}[1]{\bm{#1}}               
\newcommand{\mat}[1]{\mathsfbfit{#1}}       
\newcommand{\vecop}[1]{\bm{\mathcal{#1}}}   
\newcommand{\matel}[1]{\begin{bmatrix} #1 \end{bmatrix}}    

\newcommand{\matO}{\mathbfsf{0}}

\newcommand{\rr}{\bm{r}}               
\newcommand{\rp}{\bm{r}'}              
\newcommand{\mLam}{\mat \Lambda}       
\newcommand{\mSig}{\mat \Sigma}        
\newcommand{\pLam}{\mat P^{\Lambda}}   
\newcommand{\pSig}{\mat P^{\Sigma}}    
\newcommand{\mDls}{\mat D_{\mathrm{LS}}}
\newcommand{\mZls}{\mat Z_{\mathrm{LS}}}

\newcommand{\n}{\hat{\bm{n}}}               

\newcommand{\dd}{\mathrm{d}}  

\newcommand{\im}{\mathrm{j}}  

\newcommand{\R}{\mathbb{R}}

\DeclareMathOperator{\ee}{e}

\DeclareMathOperator{\diag}{diag}

\newcommand{\T}{\mr{T}}

\usepackage[utf8]{inputenc}

\usepackage[T1]{fontenc}
\usepackage{subcaption}
\usepackage{float}
\usepackage{tikz}
\usepackage{pgfplots}
\usepackage{pgfplotstable}
\usepackage{siunitx}
\usepackage{cleveref}
\pgfplotsset{compat=newest}
\pgfplotsset{every mark/.append style={solid}}
\usepgfplotslibrary{units}

\usepackage{amsthm}
\DeclareMathAlphabet{\mathbfsf}{\encodingdefault}{\sfdefault}{bx}{n}
\usepackage{bm}
\usepackage[OMLmathsfit]{isomath}

\def\BibTeX{{\rm B\kern-.05em{\sc i\kern-.025em b}\kern-.08em
    T\kern-.1667em\lower.7ex\hbox{E}\kern-.125emX}}
\begin{document}

\title{A Quasi-Helmholtz Projector Stabilized Full Wave Solver Encompassing the Eddy Current Regime}

\author{\IEEEauthorblockN{Tiffany L. Chhim*, John E. Ortiz, Lyes Rahmouni, Adrien Merlini, and Francesco P. Andriulli}
\IEEEauthorblockA{Department of Electronics and Telecommunications, Politecnico di Torino, Turin, Italy \\
tiffany.chhim@polito.it, john.ortizguzman@polito.it, lyes.rahmouni@polito.it, \\ adrien.merlini@polito.it, francesco.andriulli@polito.it}
}

\maketitle

\begin{abstract}
Despite its several qualities, the Poggio-Miller-Chang-Harrington-Wu-Tsai (PMCHWT) formulation for simulating scattering by dielectric media suffers from numerical instabilities and severe
ill-conditioning at low frequencies. While this drawback has been the object of numerous
solution attempts in the standard low-frequency breakdown regime for scattering problems, the eddy-current regime requires a specific treatment.
In this contribution, we present
an extension of the recently introduced quasi-Helmholtz projectors based preconditioning of the PMCHWT to obtain an
equation stable at low frequencies and specifically in three regimes relevant for eddy currents analyses:
(i) when the frequency decreases with a constant conductivity, (ii) when the
conductivity increases at fixed frequency and (iii) when the frequency decreases while keeping the product of the frequency and the conductivity constant. Being based on quasi-Helmholtz projectors our new strategy does not further degrade the conditioning of the original equation and is compatible with existing fast solvers.
The resulting full-wave formulation is capable of smoothly transitioning from simulations at high frequencies to the different low-frequency regimes which encompass, in particular, eddy currents applications.
Numerical results demonstrate the validity of our approach in all the regimes with a special emphasis given to eddy currents.
\end{abstract}


\section{Introduction}

One of the main approaches to solve electromagnetic scattering and radiation problems involving dielectrics via the boundary element method (BEM) leverages on the Poggio-Miller-Chang-Harrington-Wu-Tsai (PMCHWT) integral equation \cite{poggio1970integral}. This formulation is based on the electric and magnetic field integral operators (EFIO and MFIO) and therefore inherits their low-frequency problems. Consequently, the conditioning of its BEM discretization grows unbounded when the frequency decreases. This unfortunate shortcoming is traditionally remedied through Loop-Star/Loop-Tree techniques which, however, come at the cost of a deteriorated condition number \cite{andriulli2012loop}.

More recently the quasi-Helmholtz projectors \cite{andriulli2012loop,andriulli2013well} have been introduced to perform a frequency-regularizing decomposition while maintaining the original dense discretization conditioning of the formulation they are applied to. Notably, they have been successfully leveraged in the context of perfect electric conductors \cite{andriulli2013well} and penetrable objects \cite{beghein2017low}.
This stabilization was, however, only performed in the low-frequency breakdown regime of electromagnetic scattering (evanescent frequency), and is not directly applicable to the case of lossy conductors with increasingly high conductivities as in the case of Eddy Currents (EC) problems. These scenarios are central to a variety of applications, including non-destructive testing and industrial simulation of printed circuit boards. EC models can be obtained as an approximation of Maxwell's equations by neglecting the displacement currents, which is valid under the conditions $\omega \epsilon_0 / \sigma \ll 1$ and $L \omega \sqrt{\mu_0 \epsilon_0} \ll 1$, where $L$ is the diameter of the object and $\sigma$ is the conductivity \cite{hiptmair2007boundary,dirks1996quasi}. A quite remarkable contribution has recently proved the suitability and convergence of the PMCHWT solution to the eddy current one at low frequencies\cite{bonnet2018eddy}. 

It is then of particular interest to investigate stabilization strategies for the PMCHWT at low frequency and in this contribution we will regularize it in three different regimes determined by the above-described constraints: a regime where (i) the frequency decreases with a constant conductivity, (ii) the conductivity increases with a fixed frequency and (iii) both frequency and resistivity are simultaneously decreasing at the same rate. This third regime was defined to show that Maxwell's equations correspond to the EC model in the low-frequency and high-conductivity limit.
More specifically, in this contribution we extend the applicability of the projector enhanced PMCHWT to lossy conductive media via a careful analysis and tuning of the block operatorial scalings.
The resulting formulation is capable of handling both high-frequency simulations and the low-frequency regimes (i)-(iii). In particular, our new scheme can smoothly transition from high-frequency to eddy currents modeling without loss of accuracy or stability.
Numerical results are presented to confirm the validity of the proposed solution in terms of condition number and accuracy.

\section{Background and Notation}

Let $\Omega_i \subset \R^3$ be a dielectric object with boundary $\Gamma = \partial \Omega_i$ and outward normal $\n$ living in the exterior medium $\Omega_o = \R^3 \setminus \Omega_i$. The scatterer is characterized by a conductivity $\sigma$, permeability $\mu_i = \mu_o \mu_r$, and permittivity $\epsilon_i = \epsilon_o \epsilon_r$, where the $o$ denote the corresponding quantities of the outside medium. 
In the case of a lossy conductor, $\epsilon_i$ is defined as $\epsilon_i = \epsilon_o \epsilon_{rd} - \im \, \sigma / \omega$, where $\omega$ is the angular frequency of the time-harmonic impinging electromagnetic field ($\veg E^i$, $\veg H^i$) and $\epsilon_{rd} \in \R$.
The PMCHWT formulation, which relates the incident field and the equivalent surface electric and magnetic current densities $\veg j_s$ and $\veg m_s$ induced on $\Gamma$, is composed of the EFIO and MFIO, respectively defined as
\begin{align}
    \vecop T_k &= - \im k \vecop T_{A,k} + (\im k)^{-1} \vecop T_{\Phi,k} \, , \\
    \vecop T_{A,k} \veg f &= \n \times \int_{\Gamma} G_k(\rr, \rp) \, \veg f(\rp) \, \dd\rp \, , \\
    \vecop T_{\Phi,k} \veg f &= \n \times \nabla \int_{\Gamma} G_k(\rr, \rp) \, \nabla' \cdot \veg f(\rp) \, \dd\rp \, , \\
    \vecop K_k \veg f &= - \n \times \int_{\Gamma} \nabla G_k (\rr, \rp) \times \veg f(\rp) \, \dd\rp \, ,
\end{align}
where $k$ is the wave number and the Green's function is $G_k(\rr, \rp) = \frac{\ee^{- \im k |\rr - \rp|}}{4 \pi |\rr - \rp|}$.
The PMCHWT equation system then reads
\begin{equation}
    \begin{bmatrix}
        \eta_o \vecop T_{k_o} + \eta_i \vecop T_{k_i} & - ( \vecop K_{k_o} + \vecop K_{k_i} ) \\
        \vecop K_{k_o} + \vecop K_{k_i} & \frac{1}{\eta_o} \vecop T_{k_o} + \frac{1}{\eta_i} \vecop T_{k_i}
    \end{bmatrix}
    \begin{bmatrix}
        \veg j_s \\
        \veg m_s
    \end{bmatrix}
    =
    \begin{bmatrix}
        \n \times \veg E^i \\
        \n \times \veg H^i
    \end{bmatrix} \, ,
\end{equation}
where $\eta_m = \sqrt{\mu_m / \epsilon_m}$ is the impedance of the medium and $k_m = \omega \sqrt{\mu_m  \epsilon_m}$ ($m = i, o$). To numerically solve this equation via BEM, the unknown current densities are discretized on triangular elements using the Rao-Wilton-Glisson (RWG) basis functions $\veg f_n$ ($\veg j_s = \sum_{n = 1}^{N} [\mat j]_n \veg f_n$ and $\veg m_s = \sum_{n = 1}^{N} [\mat m]_n \veg f_n$) and the resulting equations are tested with rotated RWGs ($\n \times \veg f_n$). The discretized PMCHWT system becomes
\begin{multline}
    \mat Z \mat x =
    \begin{bmatrix}
        \eta_o \mat T_{k_o} + \eta_i \mat T_{k_i} & - ( \mat K_{k_o} + \mat K_{k_i} ) \\
        \mat K_{k_o} + \mat K_{k_i} & \frac{1}{\eta_o} \mat T_{k_o} + \frac{1}{\eta_i} \mat T_{k_i}
    \end{bmatrix}
    \begin{bmatrix}
        \mat j \\
        \mat m
    \end{bmatrix}
    =
    \begin{bmatrix}
        \mat e \\
        \mat h
    \end{bmatrix} \, ,
\end{multline}
where $[\mat T_k]_{mn} = \langle \n \times \veg f_m , \vecop T_k \veg f_n \rangle$, $[\mat K_k]_{mn} = \langle \n \times \veg f_m , \vecop K_k \veg f_n \rangle$, $[\mat e]_{m} = \langle \n \times \veg f_m , \n \times \veg E^i \rangle$ and $[\mat h]_{m} = \langle \n \times \veg f_m , \n \times \veg H^i \rangle$, with $\langle . , . \rangle$ denoting the $L_2(\Gamma)$ inner product.

Given that it includes the EFIO, it is evident that the PMCHWT equation is subject to a low-frequency breakdown, which causes the system to become severely ill-conditioned at low frequencies. This ill-conditioning can be cured by separating and independently rescaling the solenoidal and non-solenoidal parts of the equation, via e.g. Loop/Star techniques. The reader may refer to \cite{andriulli2013well} for more details on the obtention of the RWG to loop and RWG to star transformation matrices $\mLam$ and $\mSig$. This decomposition can also be achieved with the recently introduced quasi-Helmholtz projectors, which do not degrade the conditioning of the system to which they are applied, unlike Loop/Star. These projectors are defined from the traditional Loop and Star matrices as $\pSig = \mSig \left( \mSig^{\T} \mSig \right)^{+} \mSig^{\T}$ and $\pLam = \mLam \left( \mLam^{\T} \mLam \right)^{+} \mLam^{\T}$
where $^+$ denotes the Moore-Penrose pseudo inverse and can be computed in near linear complexity via multigrid preconditioning \cite{andriulli2013well}.

\section{Stabilization for Low-Frequency and High-Conductivity Regimes}

The condition number breakdowns occurring in regimes (i-iii) can be brought to light by a scaling analysis
of the Loop-Star-decomposed matrix $\mZls = \mDls \mat Z \mDls$ where
$\mDls = \diag({\matel{\mLam & \mSig}, \matel{\mLam & \mSig}})$, for which the block scalings
are
\begin{equation}
\mZls = \bordermatrix{      & \mLam   & \mSig & \mLam &\mSig \cr
                  \mLam & \omega  & \omega & \omega^2 & 1 \cr
                  \mSig & \omega  & \omega + \frac{1}{\omega} + \frac{1}{\sigma} & 1 & 1 \cr
                  \mLam & \omega^2 & 1 & \sigma + \omega & \sigma + \omega \cr
                  \mSig & 1       & 1 & \sigma + \omega & \sigma + \omega + \frac{1}{\omega}} \, ,
\end{equation}
where we have omitted the $\mc{O}$ notation for readability and where we have used the notable properties $\mLam^{\T} \mat T_{\Phi,k} = \matO$, $\mat T_{\Phi,k} \mLam = \matO$ and $\mLam^{\T} \mat K_k \mLam \sim \mc{O}(\omega^2)$ \cite{cools2009nullspaces}.
Blocks scaling as $1/\omega$ or $\sigma$  grow unbounded as the frequency decreases ($\omega \to 0$) and as the conductivity increases ($\sigma \to \infty$) and are the cause of the breakdowns in regimes (i) and (ii). In regime (iii) the product $\sigma \omega$ remains constant while the frequency decreases (or equivalently, the conductivity increases).


The pathological behaviors could be cured immediately by adequately scaling the blocks of $\mDls$, but this approach would further degrade the dense discretization conditioning of the equation. Instead, we build projector based decomposition operators
$\mat M_n = \alpha_{1,n} \pLam + \im \, \alpha_{2,n} \pSig$,
where $\alpha_{1,n}$ and $\alpha_{2,n}$ are carefully tailored scalings capable of preventing the problematic terms to grow unbounded in each regime, and the decomposed PMCHWT matrix is
\begin{equation}
    \mat Z' = 
    \begin{bmatrix}
        \mat M_1 & \matO \\
        \matO & \mat M_2
        
    \end{bmatrix}
    \mat Z
    \begin{bmatrix}
        \mat M_1 & \matO \\
        \matO & \mat M_2
    \end{bmatrix} \, .
\end{equation}
%
%
%
The rescaling factors for the different regimes are given by
\begin{enumerate}[(i)]
    \item $\sigma$ constant, $\omega \to 0$
    \begin{equation}
        \begin{bmatrix}
            \alpha_{1,1} & \alpha_{2,1} \\
            \alpha_{1,2} & \alpha_{2,2}
        \end{bmatrix}
        =
        \begin{bmatrix}
            \frac{1}{\sqrt{\omega}} & \sqrt{\omega} \\
            1 & \sqrt{\omega}
        \end{bmatrix} \, 
    \end{equation}

    \item $\omega$ constant, $\sigma \to \infty$
    \begin{equation}
        \begin{bmatrix}
            \alpha_{1,1} & \alpha_{2,1} \\
            \alpha_{1,2} & \alpha_{2,2}
        \end{bmatrix}
        =
        \begin{bmatrix}
            1 & 1 \\
            \frac{1}{\sqrt{\sigma}} & \frac{1}{\sqrt{\sigma}}
        \end{bmatrix} \, 
    \end{equation}
    
    \item $\sigma \omega$ constant, $\omega \to 0$
    \begin{equation}
        \begin{bmatrix}
            \alpha_{1,1} & \alpha_{2,1} \\
            \alpha_{1,2} & \alpha_{2,2}
        \end{bmatrix}
        =
        \begin{bmatrix}
            \left( \frac{\sigma}{\omega} \right)^{1/4} & \left( \frac{\omega}{\sigma} \right)^{1/4} \\
            \left( \frac{\omega}{\sigma} \right)^{1/4} & \left( \frac{\omega}{\sigma} \right)^{1/4}
        \end{bmatrix} \, 
    \end{equation}

\end{enumerate}
These rescalings can be shown to stabilize the conditioning of the system for each regime, in addition the projectors should be multiplied with additional diagonal preconditioning factors to further reduce the stabilized condition number. We omit here the technical underpinnings technique for the sake of brevity.

\section{Numerical Results}

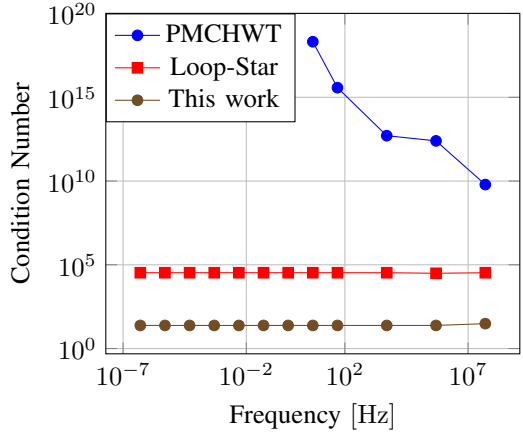
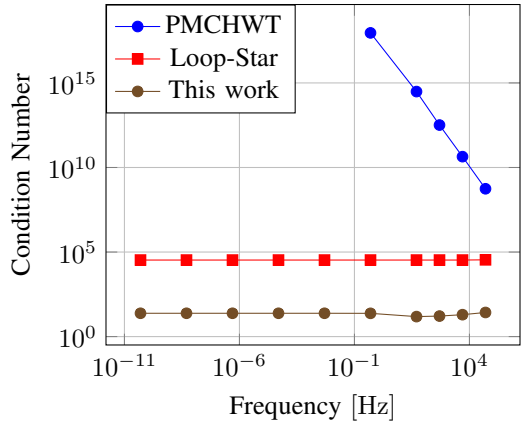
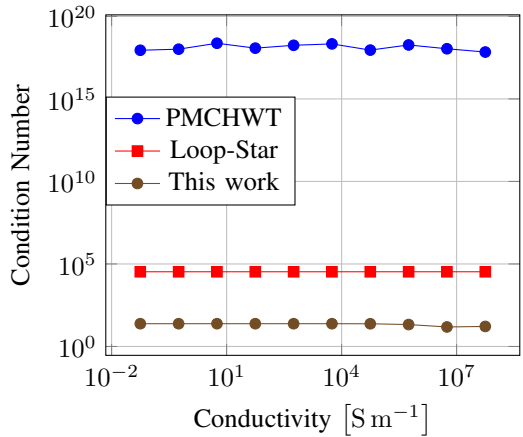
\begin{figure}
\centering
    \begin{subfigure}{0.8\columnwidth}
                \begin{tikzpicture}
       \begin{axis}[
         width=\columnwidth,
         grid=both,
         grid style={, draw=gray!10},
         major grid style={,draw=gray!50},
         xlabel=Frequency,
         ylabel=Condition Number,
         xmode=log,
         ymode=log,
         xtick={1e-7,1e-2,1e2,1e7},
         x unit = \si{\hertz},
         ytick={1e0, 1e5, 1e10, 1e15, 1e20},
         minor y tick num=10,
         legend style={at={(0.0,1)}, anchor=north west},
       ]
       \addplot
       table[x=freq,y=cnpm,col sep=comma] {Figures/reg1.csv};
       \addplot
       table[x=freq,y=cnls,col sep=comma] {Figures/reg1.csv};
       \addplot
       table[x=freq,y=cnqh,col sep=comma] {Figures/reg1.csv};
       \legend{PMCHWT, Loop-Star, This work}
       \end{axis}
    \end{tikzpicture}
        \caption{Regime (i): $\omega \sigma = 56 \pi$}
        \label{fig:cn1}
    \end{subfigure} 
    \par\bigskip
    \begin{subfigure}{0.8\columnwidth}
                \begin{tikzpicture}
       \begin{axis}[
         width=\columnwidth,
         grid=both,
         grid style={, draw=gray!10},
         major grid style={,draw=gray!50},
         xlabel=Frequency,
         ylabel=Condition Number,
         xmode=log,
         ymode=log,
         xtick={1e-11,1e-6,1e-1,1e4},
         x unit = \si{\hertz},
         ytick={1e0, 1e5, 1e10, 1e15, 1e20},
         minor y tick num=10,
         legend style={at={(0,1)}, anchor=north west},
       ]
       \addplot
       table[x=freq,y=cnpm,col sep=comma] {Figures/reg2.csv};
       \addplot
       table[x=freq,y=cnls,col sep=comma] {Figures/reg2.csv};
       \addplot
       table[x=freq,y=cnqh,col sep=comma] {Figures/reg2.csv};
       \legend{PMCHWT, Loop-Star, This work}
       \end{axis}
    \end{tikzpicture}
        \caption{Regime (ii): $\sigma = \SI{5.6e4}{\siemens\per\meter}$}
        \label{fig:cn2}
    \end{subfigure}
    \par\bigskip
    \begin{subfigure}{0.8\columnwidth}
                \begin{tikzpicture}
       \begin{axis}[
         width=\columnwidth,
         grid=both,
         grid style={, draw=gray!10},
         major grid style={,draw=gray!50},
         xlabel=Conductivity,
         ylabel=Condition Number,
         xmode=log,
         ymode=log,
         xtick={1e-2,1e1,1e4,1e7},
         x unit = \si{\siemens\per\meter},
         ytick={1e0, 1e5, 1e10, 1e15, 1e20},
         minor y tick num=10,
         legend style={at={(0,0.6)}, anchor=west},
       ]
       \addplot
       table[x=freq,y=cnpm,col sep=comma] {Figures/reg3.csv};
       \addplot
       table[x=freq,y=cnls,col sep=comma] {Figures/reg3.csv};
       \addplot
       table[x=freq,y=cnqh,col sep=comma] {Figures/reg3.csv};
       \legend{PMCHWT, Loop-Star, This work}
       \end{axis}
    \end{tikzpicture}
        \caption{Regime (iii): $\omega = \pi$ \si{\hertz}}
        \label{fig:cn3}
    \end{subfigure}
    \caption{Condition number of the formulations in regimes (i-iii) on a sphere of radius \SI{1}{\meter} discretized with \num{1048} triangles; the condition number of the PMCHWT in Figure \ref{fig:cn3} has saturated but is left for reference.}
    \label{fig:cn}
\end{figure}

The stability of the new formulations is validated through a series of numerical experiments using a sphere of radius \SI{1}{\meter} discretized with \num{1048} triangular elements. In all regimes, the standard PMCHWT formulation exhibits an extremely high condition number which saturates to machine precision. In contrast, the conditioning of our projector-based formulation remains stable in all regimes and, as expected, is much lower than those of the Loop-Star method (Figures \ref{fig:cn1} to \ref{fig:cn3}).

\begin{figure}
\centering
        \begin{tikzpicture}
   \begin{axis}[
     width=0.8\columnwidth,
     grid=both,
     grid style={, draw=gray!10},
     major grid style={,draw=gray!50},
     xlabel=Cell,
     ylabel=Magnetic Current Density,
     y unit = \si{\volt\per\meter\squared},
     minor y tick num=5,
     legend style={at={(1.00,0.0)}, anchor=south east},
   ]
   \addplot+[mark=o]
   table[x=elem,y=mana,col sep=comma] {Figures/magsol1WithEC.csv};
   \addplot+[purple,dashed,mark=x,mark options=solid]
   table[x=elem,y=mec,col sep=comma] {Figures/magsol1WithEC.csv};
   \addplot+[orange,dashdotted,mark=+,mark options=solid]
   table[x=elem,y=mqh,col sep=comma] {Figures/magsol1WithEC.csv};
   
   \legend{Analytic, EC BEM, This work}
   \end{axis}
\end{tikzpicture}
    \caption{Magnetic current density in regime (iii) ($\omega \sigma = 56 \pi$, $\omega/ 2 \pi = \SI{1e-5}{\radian\per\second}$)}
    \label{fig:magsol}
\end{figure}
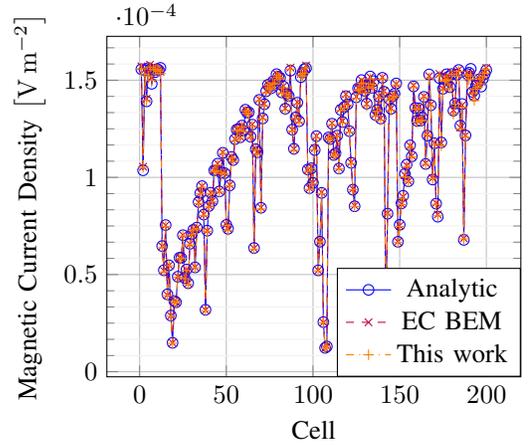

The stability of the solutions of our new formulations has been verified by comparing them to corresponding analytical results \cite{nagel2017induced}, as well as an EC model implemented from \cite{rucker1995various}. Because of space constraints we only report the results obtained in regime (iii), for which we obtain a good agreement with both analytical and eddy current solutions (Figure \ref{fig:magsol}). 

\section*{Acknowledgment}

This work was supported by the European Research Council (ERC) under the European Union’s Horizon 2020 research and innovation programme (grant agreement No 724846, project 321).

\bibliographystyle{IEEEtran}
\bibliography{references}

\end{document}